\documentclass{elsart}
\usepackage{graphicx}

\begin{document}

\begin{frontmatter}

\title{Non-Poisson distribution of
the time distances between two consecutive clusters of earthquakes} 
\author[difi]{Luigi Palatella},
\author[ilc]{Paolo Allegrini},
\author[difi,ipcf,unt]{Paolo Grigolini}, 
\author[ct]{Vito Latora},
\author[difi]{Mirko S. Mega}, 
\author[ct]{Andrea Rapisarda},
\author[ct]{Sergio Vinciguerra}
\address[difi]{INFM - Dipartimento di
Fisica dell'Universit\`a di Pisa\\ via Buonarroti 2, 56127 Pisa,
Italy} 
\address[ilc]{Istituto di Linguistica Computazionale del CNR, \\ Via
Moruzzi 1, 56124, Pisa, Italy} 
\address[ipcf]{Istituto dei Processi Chimico Fisici del CNR,\\
Via G. Moruzzi 1,56124 Pisa, Italy} 
\address[unt]{Center for Nonlinear
Science, University of North Texas,P.O. Box 311427, Denton, Texas
76203-1427 } 
\address[ct]{Dipartimento di Fisica e Astronomia,
Universit\`a di Catania, and INFN, \\ Via S. Sofia 64, 95123 Catania,
Italy}

\maketitle

\begin{abstract}
With the help of the Diffusion Entropy technique we show the  
non-Poisson statistics of the distances between consecutive Omori's swarms
of earthquakes. We give an analytical proof of the numerical
results of an earlier paper [Mega et al., Phys. Rev. Lett. 90 (2003) 188501].
\end{abstract}


\end{frontmatter}

The model usually adopted to describe the time distribution of
earthquakes is the Generalized Poisson (GP) model
\cite{shlien,gardner,gasperini,console,godano}.  The GP model assumes
that the earthquakes are grouped into temporal clusters of events and
these {\it clusters are uncorrelated}, and, therefore, completely {\em
unpredictable}: The clusters are supposed to be distributed at random
in time and therefore the time intervals between one cluster and the
next one follow a Poisson distribution. The intra-cluster earthquakes
are in fact {\em correlated} as it is expressed by the Omori's law
\cite{omori,utsu}, an empirical law stating that the main shock,
i.e. the highest magnitude earthquake of the cluster, occurring at
time $t_{0}$ is followed by a swarm of triggered earthquakes (after
shocks) whose number (or frequency) $n(t)$ decays in time as a power
law, $n(t) \propto (t-t_{0})^{-p}$, with the exponent $p$ being very
close to $1$. If we denote with the symbol $\tau$ the time intervals
between one earthquake and the next, then right after a main shock, a
short value of $\tau$ is followed with a large probability by another
short value. For the same reason, far from a main shock and prior to
the next one, a long value of $\tau$ is followed by another long value
of $\tau$. This implies that the correlation function $\langle
(\tau_{i} - \langle\tau\rangle)(\tau_{j} - \langle\tau\rangle)
\rangle$ is not zero for $i \neq j$ and that it survives for all pairs
of seismic events in between two consecutive unpredictable
shocks. Omori's law also implies \cite{bak} that the distribution of
$\tau$, is a power law $\psi(\tau) \propto \tau^{-p}$. This espression
is valid in the time regime inside a swarm, and it is then truncated
by a sharp cutoff caused by the arrival of the next swarm.

The catalog we have studied covers the period 1976-2002 in the region
of Southern California spanning $20^0$ N -$45^0$ N latitude and
100$^{0}$ W 125$^{0}$ W longitude \cite{scsn}.  This region is crossed
by the most seismogenetic part of the San Andrea fault, which
accommodates by displacement the primarily strike-slip motion between
the North America and the Pacific plates, producing velocities up to
$47$ mm/yr \cite{turcotte}. The total number of recorded earthquakes
in the catalog is $383687$.

Herein we disprove the GP model, providing evidence for the non-Poisson
statistics of inter-cluster times, by applying to the mentioned catalog
the Diffusion Entropy (DE) technique \cite{giacomo}. Here we discuss
with analytical arguments some issues that in an earlier paper
\cite{earlier} we have examined by means of a numerical treatment.
As in the GP model, we assume that each cluster starts with an
unpredictable triggering event (it may or may not be the
main-shock). The distance between the $i$-th and the $i+1$-th cluster
is therefore the time distance between such events, which we indicate
as $\tau_{i}^{[m]}$, obeying the non-correlation property, $\langle
\tau_{i}^{[m]} \tau_{j}^{[m]} \rangle = \langle \tau^{[m]} \rangle^2 $
if $i \neq j$. The superscript $m$ stands for main-shock, but we
actually need not to make the assumption that the triggering event is
a large earthquake. As we shall see, the DE measures statistical
properties of events with no need of identifying them.

Let us recall the definition of the DE functional
$S(t)$ as the Shannon entropy of $p(y,t)$, the probability
distribution to observe a fixed number of seismic events $y$ in a
given time interval \cite{giacomo}. This observation is equivalent to
observing the spreading of a number of walkers making one step forward
at each time where an event is met. Hence the term ``Diffusion
Entropy''.  Different trajectories are chosen with the usual method of
observing different time windows in the sequence, which is herein
assumed to be stationary. 
Let us assume that $p(y,t)$ follows the scaling law with the form
\begin{equation}
p(y,t)=\frac{1}{t^{\delta}} F \left ( \frac{y}{t^{\delta}}\right),
\label{scalinglaw}
\end{equation}
where $\delta$ is a positive exponent and $F(x)$ is a positive and
integrable function of x.
As a consequence of this assumption, after a straightforward algebra,
we find that
\begin{equation}
   S(t) = A + \delta ~ \ln (t).
   \label{keyrelation}
\end{equation}                  
This means that the entropy of the diffusion process is a linear 
function of $\ln (t)$ and a measure of the slope is equivalent to the 
determination of the scaling parameter $\delta$.
 
We now show that, in the earthquake series under investigation,
the principal source of entropy increase is 
given by the occurrence of the cluster-initiating seismic events.
Let us indicate with $\phi(\tau^{[m]})$ the probability density function
(pdf) of times 
between clusters and
with $h(x)$ the pdf of the number of earthquakes in a cluster.
The function $h(x)$, usually referred to as the Pareto's law of
earthquakes, is known to decay as $h(x)\simeq 1/x^{\alpha+1}$,
where $\alpha$ is a positive number.
In the literature an exponent for clusters size
distribution $\alpha$ ranging from $\alpha=1.25$ \cite{agnese} 
to $3$ \cite{console} is reported.
On the other hand, we assume for $\phi(\tau^{[m]})$, which is exponential
in the GP model, a form $\phi(\tau^{[m]}) \simeq (1/\tau^{[m]})^{\mu}$.
We also assume that $2 \leq \mu <3$. As it will become clear in the
next paragraphs, beyond the upper limit it is impossible for the DE
to distinguish between an inverse-power law and an exponential;
for $\mu<2$, on the other hand, the signal cannot be stationary 
\cite{giacomo}.
The connection between $x$ and the early defined variable $y$,
over which the DE is calculated, is
\begin{equation}
y(t)=\sum\limits_{i=1}^{z(t)} x_i,
\end{equation}
where the sum is carried over different clusters, $x_i$ is the
number of shocks in the $i$-th cluster and $z(t)$ is the number of 
clusters in the same time window of length $t$ considered for $y$.

Let us call $\widehat{\phi}(s)$ the Laplace transform of
$\phi(\tau^{[m]})$ and $\widehat{h}(k)$ the Fourier transform of
$h(x)$. 
Suppose now that the time duration of a
cluster is negligible with respect to the mean time distance
$\langle \tau^{[m]} \rangle$. In this case we can use directly a
Continuous Time Random Walk (CTRW) formalism to calculate the
probability $p(y,t)$ that a random walker, moving of a quantity
$x$ at the time at which there is a cluster of size $x$ (and
resting otherwise), is at position $y$ after a time $t$. For the
Fourier-Laplace transform of $p(y,t)$ we will have, from the theory of
CTRW \cite{Montroll,masoliver},
\begin{equation}\label{ctrw}
\widehat{p}(k,s) = \frac{1-\widehat{\phi}(s)}{s}
\frac{1}{1-\widehat{\phi}(s)\widehat{h}(k)}.
\end{equation}
To obtain the asymptotical behavior we write,
if $2<\mu<3$,
\begin{equation}
\widehat{\phi}(s) \simeq 1 - \langle \tau^{[m]} \rangle s + c
s^{\gamma},
\end{equation}
where $\gamma \equiv \mu-1$. Eq.(\ref{ctrw}) becomes
\begin{equation}\label{prima}
\widehat{p}(k,s) \simeq \frac{\langle \tau^{[m]} \rangle}{1 -
\widehat{h}(k) + s\langle \tau^{[m]} \rangle \widehat{h}(k) - c
\widehat{h}(k) s^{\gamma}}.
\end{equation}
Considering that
\begin{equation}
\widehat{h}(k) \simeq 1 + i k \langle x \rangle +  b k^{\alpha},
\end{equation}
 if $1<\alpha<2$,
we see that Eq. (\ref{prima})
leads to a ballistic scaling in the laboratory reference
frame. The DE is insensitive to drifts, so we assume that
Eq. (\ref{scalinglaw}) is fulfilled 
in a ``detrended'' moving reference frame, namely, 
where the position of the walker $y$ fulfills the condition
$\langle y(t) \rangle =0$ at each time $t$. To obtain
$\widehat{p}(k,s)$ in such a moving reference frame we have to
perform the following substitution
\begin{equation}\label{trasla}
s \rightarrow s + i k \frac{\langle x \rangle}{\langle \tau^{[m]}
\rangle}.
\end{equation}
After that, we apply the diffusive limit $k \gg s$ obtaining
\begin{equation}\label{grande}
\widehat{p}(k,s) \simeq \frac{\langle \tau^{[m]} \rangle}{ s \langle
\tau^{[m]} \rangle - b k^{\alpha} - c \left ( {i\langle x \rangle
}/{\langle \tau^{[m]} \rangle} \right )^{\gamma} k^{\gamma}},
\end{equation}
thus proving that the most anomalous, namely, the smallest between the
two exponents, either $\alpha + 1$ or $\mu$, determines the asymptotic
scaling according to the prescription $\delta = 1/\alpha$ or $\delta = 1/(\mu
-1)$, respectively.

\begin{figure}
\begin{center}
\includegraphics[angle=0,width=10.2 cm]{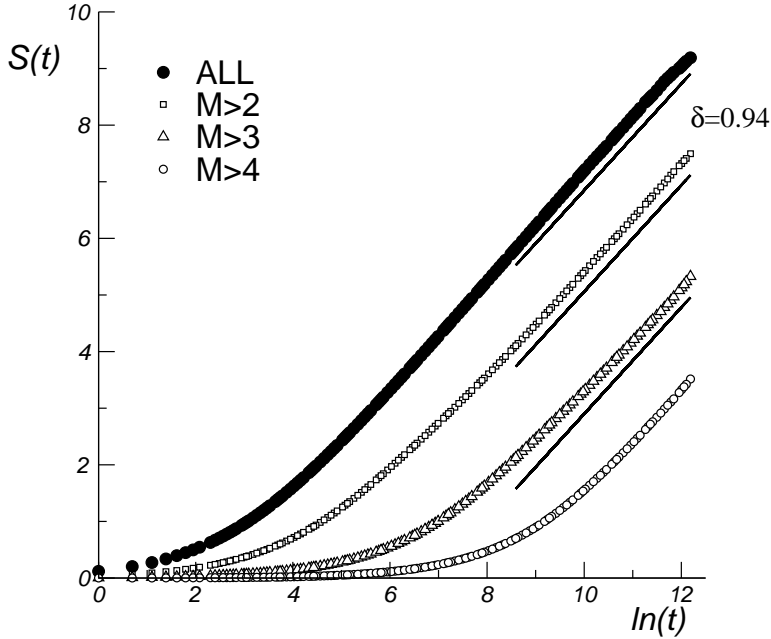}
\caption{\label{fig2} 
The Shannon entropy $S(t)$ of the diffusion process as a function of
the logarithm of time, expressed in minutes.  From top to bottom, the
curves refer to all events (full circles) and to events with threshold
$\bar M = 2,3,4$ (open symbols). The straight lines have the slope
$\delta = 0.94$.}
\end{center}
\end{figure}


In Fig. \ref{fig2} we report the results of the DE method. 
In full circles we plot the entropy $S(t)$ as a function of time 
when all the seismic events of the catalog are considered  
(independently of their magnitude $M$). 
A fit in the linear region gives a value of the scaling parameter 
$\delta = 0.94$.
We next consider (open symbols in Fig. \ref{fig2}) only 
the earthquakes with magnitude larger than a fixed value 
$\bar M = 2, 3, 4$. 
We see that, regardless of the value of the threshold $\bar M$
adopted, the function $S(t)$ is characterized by the same long-time
behavior with the same slope.  
This indicates that we are observing 
a property of the time location of the main earthquakes.
Moreover, if the value of $\delta$ were due to $h(x)$ 
we should observe at most $\delta=1/\alpha=1/1.25=0.80$.
This leads us to conclude that 
the asymptotic form (and scaling) of $p(y,t)$ is determined by the
probability $p(z,t)$ of finding $z$ unpredictable events in a time
window of duration $t$. This is in turn determined by a
$\phi(\tau^{[m]})$ decaying as an inverse power law with exponent $\mu
= 2.06$. Using the theory of Ref. \cite{giacomo} we also determine the form
of $p(z,t)$, which is well approximated by an asymmetric L\'evy
distribution with L\'evy index $\mu-1$.

In conclusion, in this paper we have studied the statistical properties of
earthquakes time distribution. Inter-cluster distances
obey an inverse power law prescription $\phi(\tau^{[m]}) \propto
(\tau^{[m]})^{-\mu}$ with $\mu = 2.06 \pm 0.01$ thus ruling out the GP model.
The method proposed is based on the fact that the asymptotic
properties of diffusion process generated by the seismic events are
scarsely sensitive to the memory stemming from the Omori's law. They
are, on the contrary, sensitive to the anomalous statistics generated
by the non-Poisson nature of the time distance between two consecutive
large earthquakes. This non-Poisson behavior reflects, in our opinion,
the cooperative behavior of the geological processes triggering the
main shock, and consequently, to some extent, some sort of
predictability. The emergenge of this possibility will be
investigated in the future.

As a final remark, we recall that throughout our analysis we made the
reasonable hypothesis that the series is stationary.  On a formal
ground, a scaling value $\delta \sim 1$ is compatible with a
nonstationary process. These nonstationary contributions may have
different orgins. The most trivial is the lack of statistics: we may
have for instance, a small number of Omori's swarms, but very extended
in time. For this reason we repeated the analysis using only portions of the
series, obtaining the same results.  Another, more interesting, source
of non-stationarity could be a strong correlation among the series of
the triggering processes, namely the condition $\langle \tau_{i}^{[m]}
\tau_{j}^{[m]} \rangle \neq \langle \tau^{[m]} \rangle^2 $. It is
interesting to notice that also this condition might reflect a form of
predictability to assess through a properly tailored form of
statistical analysis of time series. We leave this as a subject of
further investigation.

PG thankfully acknowledges the Army Research Office for financial
support through Grant DAAD19-02-0037.

\end{document}